\documentclass[a4paper,preprint,showpacs,amssymb,pre,superscriptaddress]
              {revtex4}
\usepackage{graphicx}

\begin{document}

\title{Bulk Mediated Surface Diffusion: Finite System Case}

\author{Jorge A. Revelli}
\affiliation{Grupo de F\'{\i}sica Estad\'{\i}stica, Centro
At\'omico Bariloche and Instituto Balseiro, \\ 8400 San Carlos de
Bariloche, Argentina}

\author{Carlos. E. Budde}
\author{Domingo Prato}
\affiliation{Facultad de Matem\'aticas, Astronom\'{\i}a y
F\'{\i}sica, Universidad Nacional de C\'ordoba
\\ 5000 C\'ordoba, Argentina}

\author{Horacio S. Wio$^{1,}$}
\affiliation{Departament de F\'{\i}sica, Universitat de les Illes
Balears and IMEDEA \\ E-07122 Palma de Mallorca, Spain}



\begin{abstract}

We address the dynamics of adsorbed molecules (a fundamental issue
in surface physics) within the framework of a Master Equation
scheme, and study the diffusion of particles in a finite cubic
lattice whose boundaries are at the $z=1$ and the $z=L$ planes
where $L = 2,3,4,...$, while the $x$ and $y$ directions are
unbounded. As we are interested in the effective diffusion process
at the interface $z = 1$, we calculate analytically the
conditional probability for finding the system on the $z=1$ plane
as well as the surface dispersion as a function of time and
compare these results with Monte Carlo simulations finding an
excellent agreement.
\end{abstract}

\pacs{05.40.Fb, 02.50.Ey, 05.10.Ln, 46.65.+g}

\maketitle

\vskip 1.truecm

\newpage

\normalsize

\section{Introduction}

The dynamics of adsorbed molecules is a fundamental issue in
interface science
\cite{c6r1,c6r2,c6r3,c6r4,c6r5,c6r6,c6r7,c6r8,c6r9,c6r10} and is
also crucial in a large number of emerging technologies
\cite{c6r3,c6r11}. Adsorption at the solid-liquid interfaces
arises, for instance, in many biological contexts involving
protein deposition \cite{c6r12,c6r13,c6r14}, in solutions or melts
of synthetic macromolecules \cite{c6r15,c6r16,c6r17,c6r18}, in
colloidal dispersions \cite{c6r19}, and in the manufacture of
self-assembly mono- and multi-layers
\cite{c6r20,c6r21,c6r22,c6r11}.

Experimental studies of surfactant molecules
\cite{c6r5,c6r6,c6r7}, proteins \cite{c6r9,c6r10,c6r24} and
synthetic polymers \cite{c6r8,c6r10} confined to interfaces have
identified different types of surface translational motions. One
is in-surface self-diffusion of individual molecules, which has
been investigated with fluorescence recovery after photobleaching
(FRAP) methods \cite{c6r4,c6r10,c6r24}. Measured self-diffusivity
at liquid-solid interfaces are much smaller than bulk values
\cite{c6r24} and similar to bulk values in liquid-gas cases
\cite{c6r10}. A second source of motion, exclusive to the
liquid-fluid interface, is surface visco-elasticity. Adsorbed
surface phases possess compressibility and viscosities which
govern the dynamics of the surface density waves. Their relaxation
kinetics have been intensively studied experimentally
\cite{c6r5}-\cite{c6r7} and theoretically
\cite{c6r1,c6r2,c6r25,c6r26}, with direct viscoelastic measurement
\cite{c6r5} and surface light scattering studies
\cite{c6r4,c6r25,c6r26} providing considerable support for current
theories.

In this paper we explore another mechanism called {\it
bulk-mediated surface diffusion}. This mechanism arises at
interfaces separating a liquid bulk phase and  a second phase
which may be either solid, liquid, or gaseous. Whenever the
adsorbed species are soluble in the liquid bulk,
adsorption-desorption processes occur continuously. These
processes generate surface displacement because desorbed molecules
undergo Fickian diffusion in the liquid's bulk, and are later
re-adsorbed elsewhere. When this process is repeated many times,
an effective diffusion results for the molecules on the surface.
The importance of bulk-surface exchange in relaxing homogeneous
surface density perturbations is experimentally well established
\cite{c6r1,c6r2,c6r5,c6r9,c6r27}.

Here, unlike our previous work \cite{c6r28} on the
adsorption-desorption mechanism, we assume that the bulk is
finite. The main goal of this work is the observation of an
effective diffusion process at the interface $z=1$. We calculate
analytically its variance, $\langle r^2(t) \rangle$, as a function
of time and $P(x,y,z=1;t|0,0,1,t=0)$, the conditional probability
of being on the surface at time $t$ since the particle arrived at
$t=0$, that we indicate as $P_{z=1}(t)$.

The organization of the paper is as follows. In the next section
we describe the model by means of a set of Master Equations and
solve this system in Fourier-Laplace space. After that, we present
a special case (the bilayer case) where we can obtain explicit
analytical results for $P_{z=1}(t)$ and $\langle r^2(t) \rangle$.
Next we compare these theoretical results with numeric Monte Carlo
simulations. Moreover, we present results obtained by numerical
inverse transformation of the equations for $\langle r^2 \rangle$
and $P_{z=1}$ for the cases with more layers. In the last section
we discuss the results  and draw some conclusions.

\section{The Model}

Let us start with the problem of a particle making a random walk
in a finite cubic lattice. The bulk is bounded in the $z$
direction where the particles can move from $z = 1$ to $z = L$.
The $x$ and $y$ directions remain unbounded. The position of the
walker is defined by a vector $\vec{r}$ whose components are
denoted by the integer numbers $n,m,l$ corresponding to the
directions $x$, $y$ and $z$ respectively.

The probability that the walker is at $(n,m,l)$ for time $t$ given
it was at $(0,0,l_0)$ at $t=0$, $P(n,m,l;t | 0,0,l_0,t=0) =
P(n,m,l;t)$, satisfies the following master equation
\begin{eqnarray}
\label{c6mod1}
\dot{P}(n,m,l;t) & = & \gamma P(n,m,l+1;t) - \delta P(n,m,l;t)  \nonumber \\
                 &   & + \alpha^{1} [P(n-1,m,l;t)+P(n+1,m,l;t)-2 P(n,m,l;t)] \nonumber \\
                 &   & + \beta^{1} [P(n,m-1,l;t)+P(n,m+1,l;t)-2 P(n,m,l;t)], ~~~~\mbox{for $l=1$} \nonumber \\
\dot{P}(n,m,l;t) & = & \alpha [P(n-1,m,l;t)+P(n+1,m,l;t)-2 P(n,m,l;t)] \nonumber \\
                 &   & + \beta [P(n,m-1,l;t)+P(n,m+1,l;t)-2 P(n,m,l;t)] \nonumber \\
                 &   & + \gamma P(n,m,l+1;t) + \delta P(n,m,l-1;t) - 2 \gamma P(n,m,l;t), ~~~~\mbox{for $l=2$} \nonumber \\
\dot{P}(n,m,l;t) & = &  \alpha [P(n-1,m,l;t)+P(n+1,m,l;t)- 2 P(n,m,l;t)] \nonumber \\
                 &   & + \beta [P(n,m-1,l;t)+P(n,m+1,l;t)-2 P(n,m,l;t)] \nonumber \\
                 &   & + \gamma [P(n,m,l+1;t) +  P(n,m,l-1;t) - 2 \gamma [P(n,m,l;t)], ~~~~\mbox{for $3 \leq l \leq L-1$} \nonumber \\
\dot{P}(n,m,l;t) & = &  \gamma P(n,m,l-1;t) - \gamma P(n,m,l;t) \nonumber \\
                 &   & + \alpha [P(n-1,m,l;t)+P(n+1,m,l;t)-2 P(n,m,l;t)] \nonumber \\
                 &   & + \beta [P(n,m-1,l;t)+P(n,m+1,l;t)-2 P(n,m,l;t)], ~~~~\mbox{for $l = L.$}
\end{eqnarray}
where $\alpha, \beta$ and $\gamma $ are the bulk transition
probabilities per unit time in the $x$,$y$ and $z$ directions
respectively, and $\delta$ is the desorption probability per unit
time from the boundary plane $z=1$.

It is important to note that the model presented in Eq.
(\ref{c6mod1}), allows the possibility that the particles can move
in the plane $z=1$ with temporal frequencies $\alpha^{1}$ in the
$x$ direction and $\beta^{1}$ in the $y$ direction. If these
temporal frequencies are equal to zero, the motion through the
$z=1$ plane is exclusively due to the dynamics across the bulk. In
addition we can observe that this is a finite set of $L$
equations. This fact establishes an important difference with the
infinite set of equations presented in  \cite{c6r28}, a crucial
difference because this generates different solutions to the
problem. In order to solve this finite set of equations, we take
the Fourier transform with respect to the $x$ and $y$ variables,
and the Laplace transform in the $t$ variable. After these
transformations, we obtain the following set of equations
\begin{eqnarray}
\label{c6mod2}
s G(k_x,k_y,l;s) - P(k_x,k_y,l,t=0) & = &  \gamma G(k_x,k_y,l+1;s) \nonumber \\
                                    &   &  - \delta G(k_x,k_y,l;s) + A^{1}(k_x,k_y) G(k_x,k_y,l;s),~~~~\mbox{for $l=1$} \nonumber \\
s G(k_x,k_y,l;s) - P(k_x,k_y,l,t=0) & = &  A(k_x,k_y) G(k_x,k_y,l;s)+ \delta G(k_x,k_y,l-1;s) \nonumber \\
                                    &   & + \gamma G(k_x,k_y,l+1;s)-2 \gamma G(k_x,k_y,l;s),~~~~~\mbox{for $l=2$} \nonumber \\
s G(k_x,k_y,l;s) - P(k_x,k_y,l,t=0) & = &  A(k_x,k_y) G(k_x,k_y,l;s)+ \gamma [G(k_x,k_y,l-1;s) \nonumber \\
                                    &   & + \gamma G(k_x,k_y,l+1;s)-2 \gamma G(k_x,k_y,l;s)], ~~~~\mbox{for $3 \leq l \leq L-1$} \nonumber \\
s G(k_x,k_y,l;s) - P(k_x,k_y,l,t=0) & = & A(k_x,k_y) G(k_x,k_y,l;s) \nonumber \\
                                    &   & - \gamma G(k_x,k_y,l;s)+  \gamma G(k_x,k_y,l-1;s), ~~~~\mbox{for $l = L,$}
\end{eqnarray}
where we have defined
\begin{eqnarray}
\label{c6def1} G(k_x,k_y,l;s) & = & G(k_x,k_y,l;s |0,0,l_0;t=0) \nonumber \\
& = & \int_{0}^{\infty} e^{-s t}
              \sum_{n,m,-\infty}^{\infty} e^{k_x n + k_y m}
               P(n,m,l;t) dt  \nonumber \\
                             & = & \mbox{} {\it L}[\sum_{n,m,-\infty}^{\infty} e^{k_x n + k_y m}
                             P(n,m,l;t)],
\end{eqnarray}
${\it L}$ indicates the Laplace transform of the quantity inside the
brackets, and
\begin{equation}
\label{c6def2} A(k_x,k_y) = 2 \alpha [\cos(k_x)-1] + 2 \beta
[\cos(k_y)-1],
\end{equation}
\begin{equation}
\label{c6def3}
 A^{1}(k_x,k_y) = 2 \alpha^{1} [\cos(k_x)-1] +
           2 \beta^{1} [\cos(k_y)-1].
\end{equation}
Using the matrix formalism, Eq. (\ref{c6mod2}) can be written as
\begin{equation}
\label{c6mat1}
[s \tilde{I} - \tilde{H}] \tilde{G} = \delta_{l l_0}= \tilde{I}_{l l_0},
\end{equation}
where $\tilde{G}$ is an $L \times L$ array that has the following
components
\begin{equation}
\label{mat2}
\tilde{G}_{l l_0} = [G[k_x,k_y,l;s | n,m,l_0;t_0 ]],
\end{equation}
$\tilde{I}$ is the identity matrix and $\tilde{H}$ is a
tri-diagonal matrix whose components are
\[
\tilde{H}=\left( \begin{array}{ccccccc}
-\delta + A^{1} &  \gamma       & 0       & 0  \cdots  &   0    &    0   & 0  \\
 \delta         &  C            & \gamma  & 0  \cdots  &   0    &    0   & 0   \\
 0              &  \gamma       & C       & \gamma     &   0    &    0   & 0   \\
 \cdot          &  \cdot        & \cdot   &    \cdots  &   0    &    0   & 0   \\
 \cdot          &  \cdot        & \cdot   &    \cdots  &   0    &    0   & 0   \\
 \cdot          &  \cdot        & \cdot   &    \cdots  &   0    &    0   & 0   \\
 0              &   0           & 0       & 0  \cdots  & \gamma &    C   & \gamma  \\
 0              &   0           & 0       & 0  \cdots  &   0    & \gamma & -\gamma + A
                   \end{array}  \right), \]
The $C$ parameter is defined as
\begin{equation}
\label{c6C}
C = -2 \gamma + A(k_x,k_y).
\end{equation}

In order to find the solution to the Eq. (\ref{c6mat1}), we
decompose the $\tilde{H}$ matrix according to
\begin{equation}
\label{c6mat4}
\tilde{H}=A(k_x,k_y) \tilde{I} +\tilde{H}_0 +
\tilde{H}_1 + \tilde{H}_2,
\end{equation}
where
\[\tilde{H}_0=\left( \begin{array}{ccccccc}
-2 \gamma &  \gamma      & 0         & 0  \cdots     &   0     &   0       &  0       \\
 \gamma   &  -2 \gamma   & \gamma    & 0  \cdots     &   0     &   0       &  0       \\
 0        &  \gamma      & -2 \gamma & \gamma \cdots &   0     &   0       &  0       \\
 \cdot    &  \cdot       & \cdot     &   \cdots      &   0     &   0       &  0       \\
 \cdot    &  \cdot       & \cdot     &   \cdots      &   0     &   0       &  0       \\
 \cdot    &  \cdot       & \cdot     &   \cdots      &   0     &   0       &  0       \\
 0        &  \cdot       & \cdot     &   \cdots      & \gamma  & -2 \gamma & \gamma   \\
 0        &  \cdot       & \cdot     &   \cdots      &   0     &   \gamma  & - \gamma
                   \end{array}  \right), \]

\[(\tilde{H}_1)_{i j}=\Delta_1 \left\{ \begin {array}{ll}
                           1      & \mbox{if i = j = 1} \\
                           0      & \mbox{otherwise}
                         \end{array}
                \right. \]

\[(\tilde{H}_2)_{i j}=\Delta_2 \left\{ \begin {array}{ll}
                           1      & \mbox{if i = 1 and j = 2} \\
                           0      & \mbox{otherwise}
                         \end{array}
                \right. \]
and
\begin{eqnarray}
\label{c6deltas}
\Delta_1 & = & -\delta - [-2 \gamma + A(k_x,k_y) - A^{1}(k_x,k_y)], \nonumber \\
\Delta_2 & = & \delta - \gamma .
\end{eqnarray}
We also define
\begin{eqnarray}
\label{c6lasG}
\tilde{G}^0 & = & [s \tilde{I} - (A(k_x,k_y) \tilde{I} + \tilde{H}_0)]^{-1}, \nonumber \\
\tilde{G}^1 & = & [s \tilde{I} - (A(k_x,k_y) \tilde{I} +
\tilde{H}_0+ \tilde{H}_1)]^{-1}.
\end{eqnarray}
A formal solution of the Eq. (\ref{c6mat1}) is
\begin{equation}
\label{c6sol}
 \tilde{G} =  [s \tilde{I}-\tilde{H}]^{-1}.
\end{equation}
We can show, by applying the Dyson formula \cite{tesis}, that
\begin{equation}
\label{c6G2}
 \tilde{G}_{l l_0} = \tilde{G}^{1}_{l l_0} + \frac{\Delta_2 \,\ \tilde{G}^{1}_{l 2} \,\ \tilde{G}^{1}_{1 l_0}}
                           {1 - \Delta_2 \,\ \tilde{G}^{1}_{1 2}},
\end{equation}

\begin{equation}
\label{c6G1}
 \tilde{G}^{1}_{l l_0} =  \tilde{G}^{0}_{l l_0} +
                          \frac{\Delta_1 \,\ \tilde{G}^{0}_{l 1} \,\ \tilde{G}^{0}_{1 l_0}}
                           {1 - \Delta_1 \,\ \tilde{G}^{0}_{1 1}}.
\end{equation}
The solution for $\tilde{G}^{0}_{l l_0}$ can be obtained
analytically \cite{c6r29}. The result is
\begin{equation}
\label{c6G0}
 \tilde{G}^{0}_{l l_0} =  \sum_{i=0}^{L-1} f_{l i} f_{l_0 i}
                          \frac{1}{2 \gamma + (s - A(k_x,k_y)) - 2 \gamma
                          \cos(q_i)},
\end{equation}
where
\begin{equation}
\label{c6f}
f_{l i} = K \sin(l q_i),
\end{equation}
and
\begin{equation}
\label{c6ql}
q_i = \frac{(2 i + 1) \pi}{2 L + 1}.
\end{equation}
The constant $K$ is obtained by exploiting the orthonormality
relations for the $f_{l i}$ functions given by
\begin{equation}
\label{c6ortonorm} \sum_{i=1} f_{i l} f_{i j} = \delta_{l j},
\end{equation}
while the completeness relation for the $f_{l i}$ functions is
written as
\begin{equation}
\label{c6comp} \sum_{i=1} f_{l i} f_{j i} = \delta_{l j}. 
\end{equation}
The corresponding expression for  $K$ is
\begin{equation}
\label{c6K} K = \frac{2}{\sqrt{2 L + 1}}.
\end{equation}

We are now able to find the statistical quantities which describe
the diffusion problem over the surface. We are interested in the
probability of finding a particle in the site $(m,n,l = 1)$ at
time $t$ given it was in $(0,0,l=1)$ at $t=0$. This quantity is
obtained by applying the inverse Laplace transform on the
$\tilde{G}_{11}$ matrix element. Another direct measurable
experimental magnitude \cite{c6r30,c6r31} is the variance
($\langle r^2(t) \rangle$) of the probability distribution at time
$t$ over the plane
\begin{equation}
\label{c6vari}
 \langle r^2(t) \rangle_{plane}.
\end{equation}
This quantity measures the particles dispersion over the surface
and has a direct relation to the diffusion coefficient. When
$P(m,n,l=1; t | 0,0,l_0=1;t=0)$ is known, the variance is
calculated in the following manner
\begin{equation}
\label{c6vari2}
\langle r^2(t) \rangle_{plane} = \sum_{m,n=-\infty}^{\infty}
P(m,n,l=1; t | 0,0,l_0=1;t=0) (m^2+n^2).
\end{equation}
In order to obtain this expression, we have assumed symmetric
properties for the diffusion along the $x$ and $y$ directions,
that is $<x(t)> = <y(t)> = 0$.

Finally the variance in the Laplace space can be obtained as
\begin{equation}
\label{variance3} <r^2(s)>_{plane} = -
[\frac{\partial^{2}}{\partial
k_{x}^{2}}+\frac{\partial^{2}}{\partial k_{y}^{2}}]
[\tilde{G}_{11}]\mid_{k_x=k_y=0}.
\end{equation}

\section{The bilayer case}

So far, we have developed a general theory describing the
important problem of the diffusion of a particle system over a
surface which is surrounded by a bounded bulk. The expressions we
found, are in the Laplace space and the obtention of the inverse
transform is usually a non trivial task.

In this section we restrict the problem to finding a particular
solution, considering the case of a system of particles moving
inside a bilayer, that is the space formed by the surface where we
investigate the movement of the particles and a second layer. To
obtain the $P(m,n,l=1; t | 0,0,l_0=1;t=0)$ ($P_{z=1}(t)$), we take
the expression in Eq. (\ref{c6sol}) with $L = 2$ and evaluate for
$k_x = k_y = 0$. Then we obtain the inverse Laplace transform. The
result is
\begin{equation}
\label{c6PZ1}
P_{z=1}(t) = \frac{\gamma}{(\gamma + \delta)} +
         \frac{\delta}{(\gamma + \delta)} \exp{[-(\gamma +
         \delta) t]}.
\end{equation}
This expression establishes that this probability is only a
function of the temporal  adsorption and desorption rates and does
not depend on the temporal rates on the plane. From this equation,
we can also obtain the initial condition of the problem by
evaluating Eq. (\ref{c6PZ1}) at $t = 0$, obtaining $P_{z=1}(t=0) =
1$.

If we consider the long time limit, the expression has the
following behavior
\begin{equation}
\label{c6tgran}
\lim_{t \to \infty} P_{z=1}(t) =
\frac{\gamma}{\gamma + \delta}.
\end{equation}
This expression shows that, at this limit,  the probability of
finding the particle in the $z=1$ plane does not decay to zero (as
happens in unbounded systems \cite{c6r28}), but it approaches an
asymptotic value which is a function of the ratio of the temporal
rates.

If the adsorption rate $\gamma$ is large (compared with the
desorption rate $\delta$), then the probability is large, the
particles go back to the plane frequently. On the other hand, if
the desorption rate is large, the particles go away from the
surface and so the probability decays. These behaviors are shown
and quantified by Eq. (\ref{c6PZ1}).

In the case of the variance, the result we have obtained can be
decomposed as follows
\begin{equation}
\label{c6vari4}
\langle r^2(t) \rangle = \langle r^2(t) \rangle_{plane} + \langle r^2(t) \rangle_{vol},
\end{equation}
where
\begin{eqnarray}
\label{c6rplano}
 \langle r^2(t) \rangle_{plane} = (\alpha^{1} + \beta^{1})
\left(\frac{-2 \gamma \delta}{(\gamma + \delta)^3}
\left(\exp{[-(\gamma + \delta) t]} - 1 \right) \right. \nonumber \\
                   \left. + \frac{2 \gamma \delta + \delta^2}{(\gamma + \delta)^2} t\, \, \exp{[-(\gamma + \delta) t]}
                    + \frac{\gamma^2}{(\gamma + \delta)^2} t
                    \right),
\end{eqnarray}
\begin{eqnarray}
\label{c6rvol}
\langle r^2(t) \rangle_{vol} = (\alpha + \beta) \left(\frac{4
\gamma \delta}{(\gamma + \delta)^3} \left(\exp{[-(\gamma + \delta) t]} - 1 \right) \right. \nonumber \\
                   \left. + \frac{2 \gamma \delta}{(\gamma + \delta)^2}\, t\, \exp{[-(\gamma + \delta) t]}
                    + \frac{2 \gamma \delta}{(\gamma + \delta)^2} t
                    \right).
\end{eqnarray}
From  Eqs. (\ref{c6vari4}), (\ref{c6rplano}) and (\ref{c6rvol}) we
can make the following observations. Firstly, the variance can be
decomposed into two contributions, one corresponding to the
particle movement across the bulk and the other to the surface
movement. The dependence of the variance on the time rates
parallel to the surface is linear, while in these same relations,
the adsorption and desorption rates enter in a more complicated
way. The functional form obtained is similar for both movements
(which however are nonindependent).

For a large evolution time, the mean square distance or dispersion
grows linear with time; each diffusive process has its own slope
or growing rate (these slopes are associated with the diffusion
coefficient). However, this is an expected behavior due to the
model we are using. The other contributions are transient ones,
that decay with a time constant $\tau = (\gamma + \delta)^{-1}$,
hence, the stronger the adsorption or desorption, the faster is
the decay of these contributions.

\section{Results}

In this section we show the theoretical results, including some
special numerical procedures, and make the comparison with Monte
Carlo simulations. In all simulations we have fixed the following
parameters: $\alpha = \beta = \gamma = 1$ and $\alpha^{1} =
\beta^{1} = 0$, and have averaged over $10^{6}$ realizations.

\begin{figure}
\centering
\resizebox{.6\columnwidth}{!}{\includegraphics{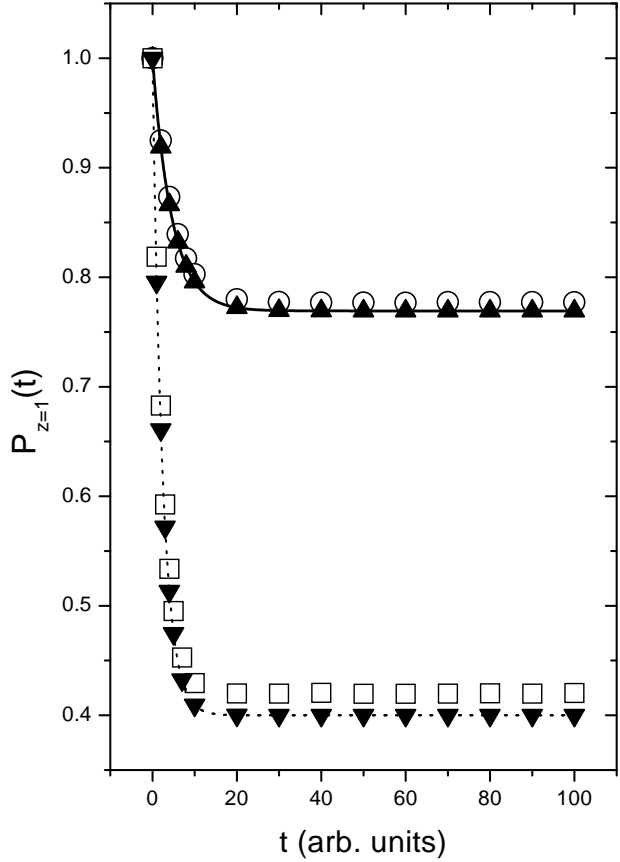}}
\caption{Temporal evolution of the $P_{z=1}(t)$. We have shown two
cases: (i) $\blacktriangle$ represent theoretical points (see Eq.
(\ref{c6PZ1})), the continuous line indicates the
theoretical-numerical results and $\bigcirc$ are the simulations
data for $\delta = 0.1$; (ii) $\blacktriangledown$ corresponds to
theoretical points, the dashed line represents
theoretical-numerical results and $\square$ the simulations data
for $\delta = 0.5$.} \label{f16}
\end{figure}

\begin{figure}
\centering
\resizebox{.6\columnwidth}{!}{\includegraphics{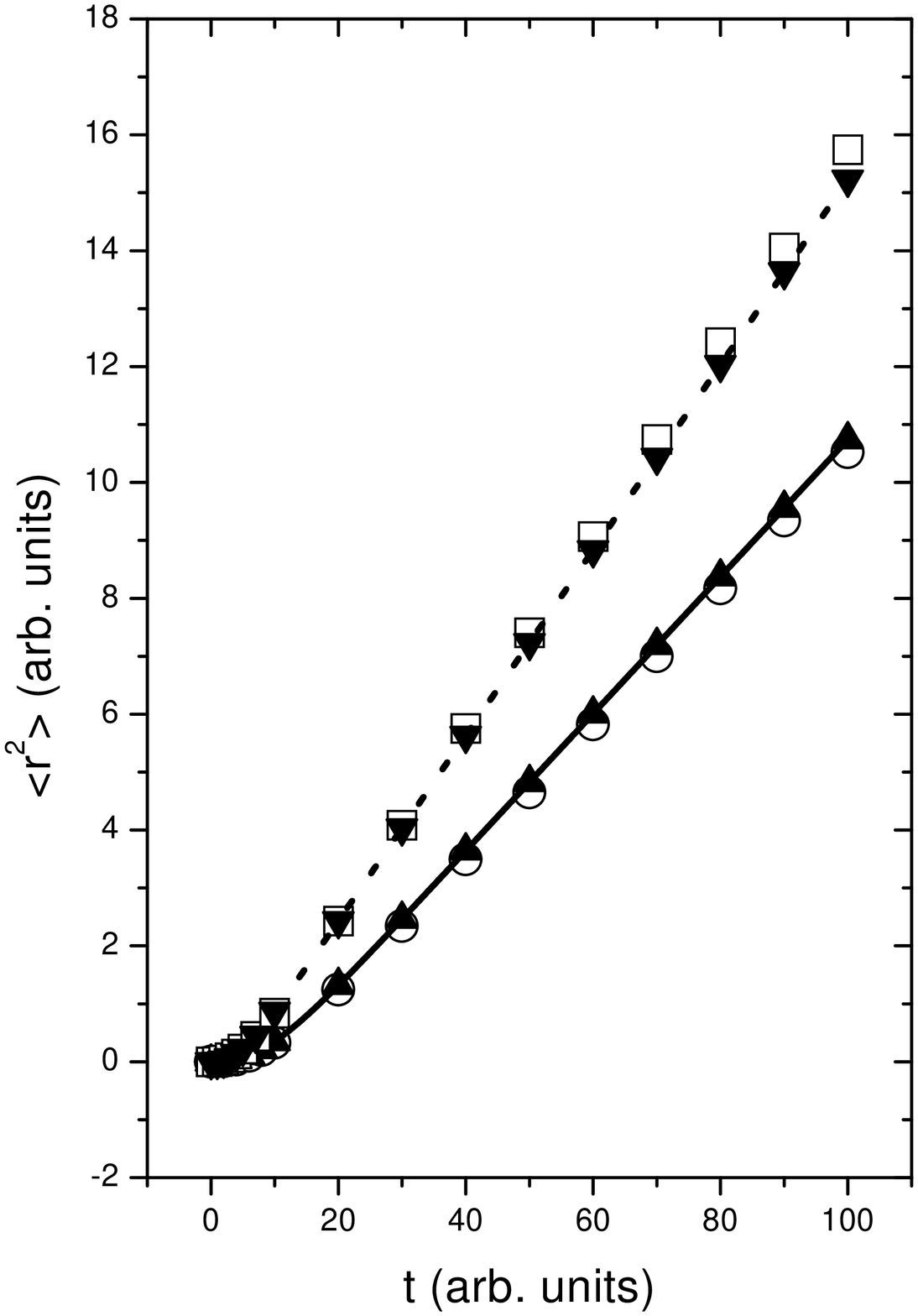}}
\caption{Time evolution of $\langle r^2 \rangle$. We have
represented two cases: (i) $\blacktriangle$ represent theoretical
points (see Eq.(\ref{c6PZ1})), the continuous line indicates the
theoretical-numerical results and $\bigcirc$ are the simulation
data for $\delta = 0.1$; (ii) $\blacktriangledown$ correspond to
theoretical points, the dashed line represents
theoretical-numerical results and $\square$ the simulations data
for $\delta = 0.5$.} \label{f26}
\end{figure}

In Fig. \ref{f16} we present the theoretical-numerical (that is
using a computer program to calculate the inverse Laplace
transform \cite{LAPIN}), theoretical results obtained from Eq.
(\ref{c6PZ1}) and simulation results for the temporal evolution of
the probability to find the system on the surface for the bilayer
case. Here we show the curves for two values of the desorption
rate $\delta$. As is apparent from the figure, there is excellent
agreement between the theoretical and simulation results. Such
excellent agreement indicates that the numerical procedure for the
obtention of the inverse Laplace transform is a reliable tool, and
that we can trust their results in those cases where analytical
results are not accessible (for instance the cases with larger
number of layers that we will consider in the following). In Fig.
\ref{f26}, and again for the bilayer case, we depict for the
variance ($\langle r^2(t) \rangle$), both theoretical and
numerical results. Again the agreement is excellent.

\begin{figure}
\centering
\resizebox{.6\columnwidth}{!}{\includegraphics{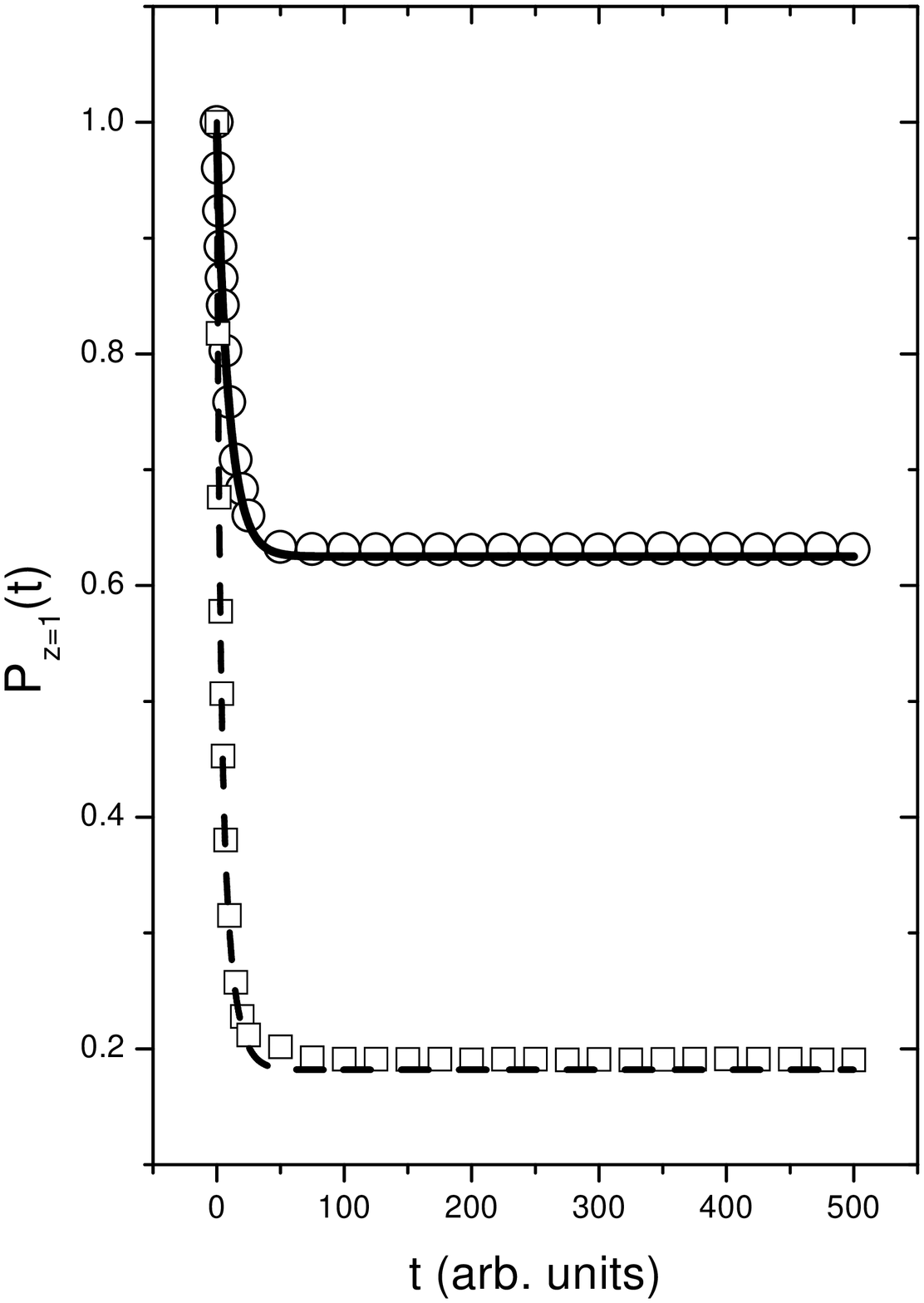}}
\caption{Time evolution of the $P_{z=1}(t)$. We have represented
two cases: (i) the continuous line depicts the theoretical result
obtained numerically and $\bigcirc$ are the simulation points for
$3$ layers; (ii) the dashed line represents theoretical results
and $\square$ the simulations for $4$ layers.} \label{f36}
\end{figure}

\begin{figure}
\centering
\resizebox{.6\columnwidth}{!}{\includegraphics{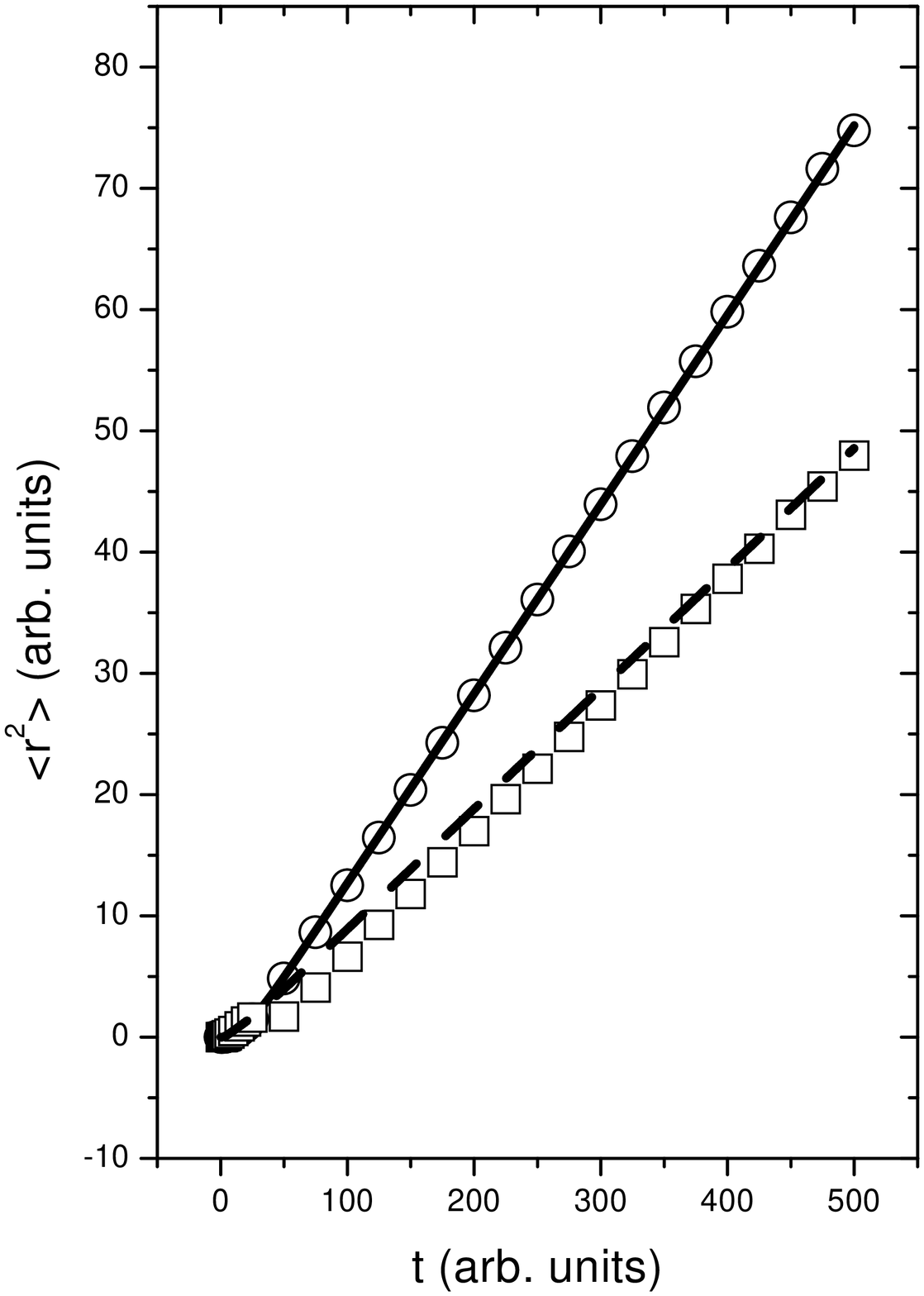}}
\caption{Time evolution of the variance $\langle r^2 \rangle$. We
have represented two cases: (i) the continuous line depicts the
theoretical result obtained numerically and $\bigcirc$ are the
simulation points for $3$ layers; (ii) the dashed line represents
theoretical results and $\square$ the simulation for $4$ layers.}
\label{f46}
\end{figure}

In Figs. \ref{f36} and \ref{f46} we present the $P(m,n,l=1; t |
0,0,l_0=1;t=0)$ and the $\langle r^2(t) \rangle$ but now for a
number of layers larger than two. Here we compare the theoretical
and numerical results again. We remark that the theoretical
results were obtained fitting the inverse Laplace transformation
of  Eqs. (\ref{c6G0}) and (\ref{c6vari3}) numerically. Again we
have found an excellent agreement between theoretical and
simulation results.

\begin{figure}
\centering
\resizebox{.6\columnwidth}{!}{\includegraphics{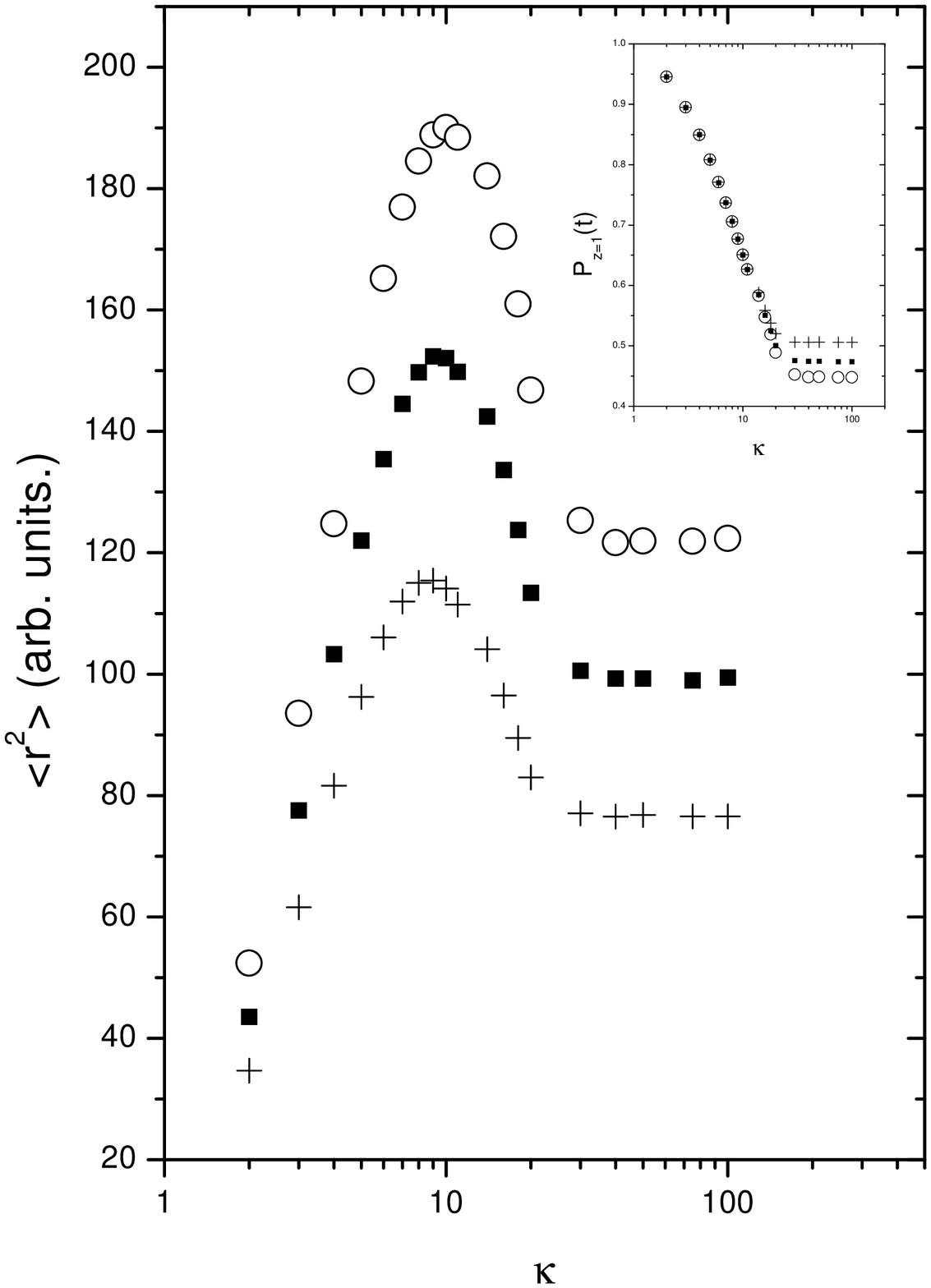}}
\caption{$\langle r^2 \rangle$ vs $\kappa$ for the case $\delta =
0.02$. The $\bigcirc$ correspond to an observation time $t=
1200$,  $\blacksquare$ is for $t=1300$ and $+$ for $t=1500$. The
insert shows the $P_{z=1}(t)$ vs $\kappa$. The system
is in the asymptotic region.} \label{f56}
\end{figure}

\begin{figure}
\centering
\resizebox{.6\columnwidth}{!}{\includegraphics{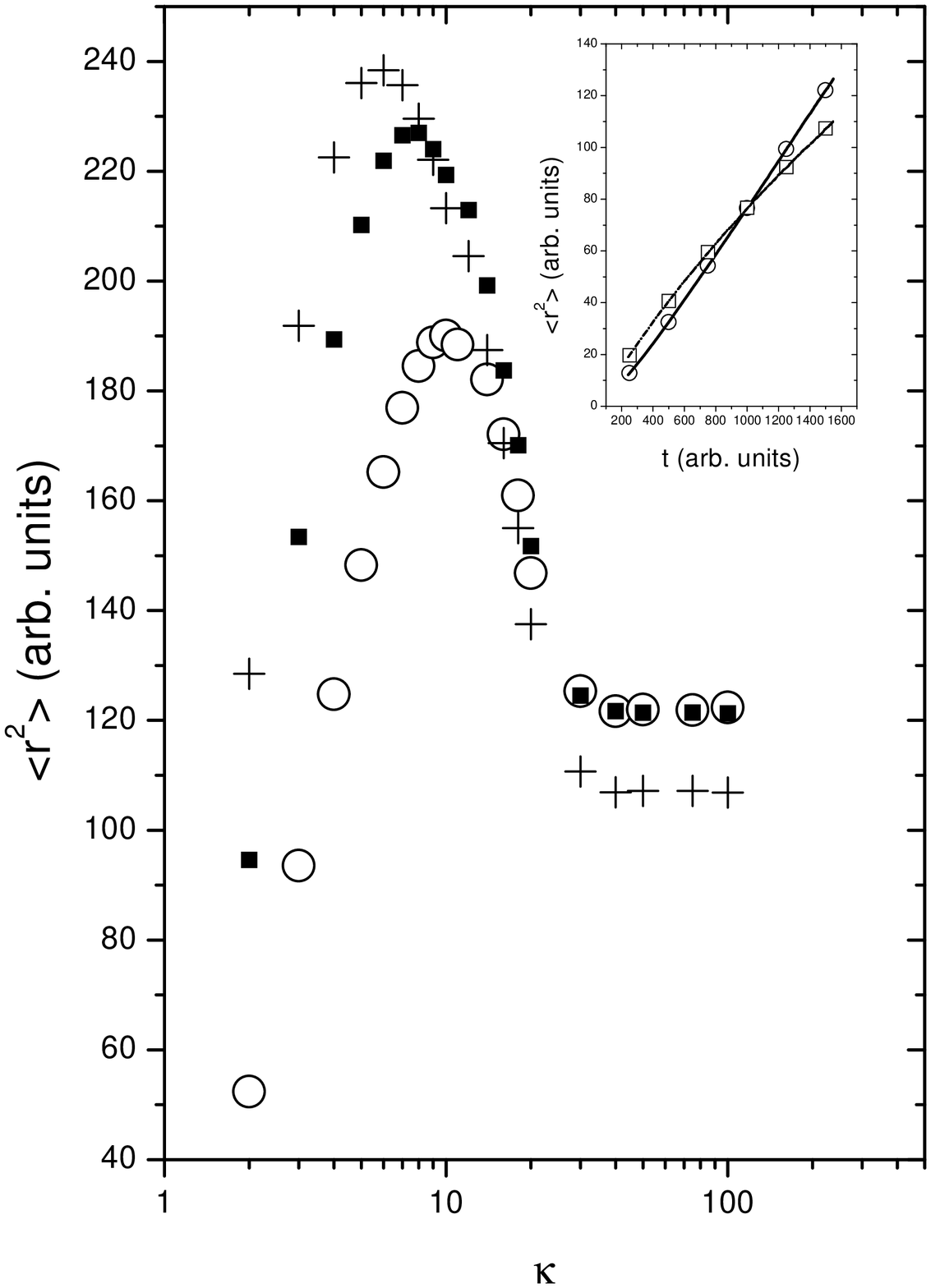}}
\caption{$\langle r^2 \rangle$ vs $\kappa$. We have represented
the case for $t = 1500$. The $\bigcirc$ correspond to a desorption
rate $\delta=0.02$, while $\blacksquare$ is for $\delta=0.04$ and
$+$ for $\delta=0.06$. Insert: time evolution of $\langle r^2
\rangle$ for $\delta = 0.02$ and $\kappa = 100$ ($\bigcirc$), and
for $\delta = 0.06$ and $\kappa = 100$ ($\Box$). The continuous
lines correspond to the infinite bulk behavior \cite{c6r28}.}
\label{f66}
\end{figure}

\begin{figure}
\centering
\resizebox{.6\columnwidth}{!}{\includegraphics{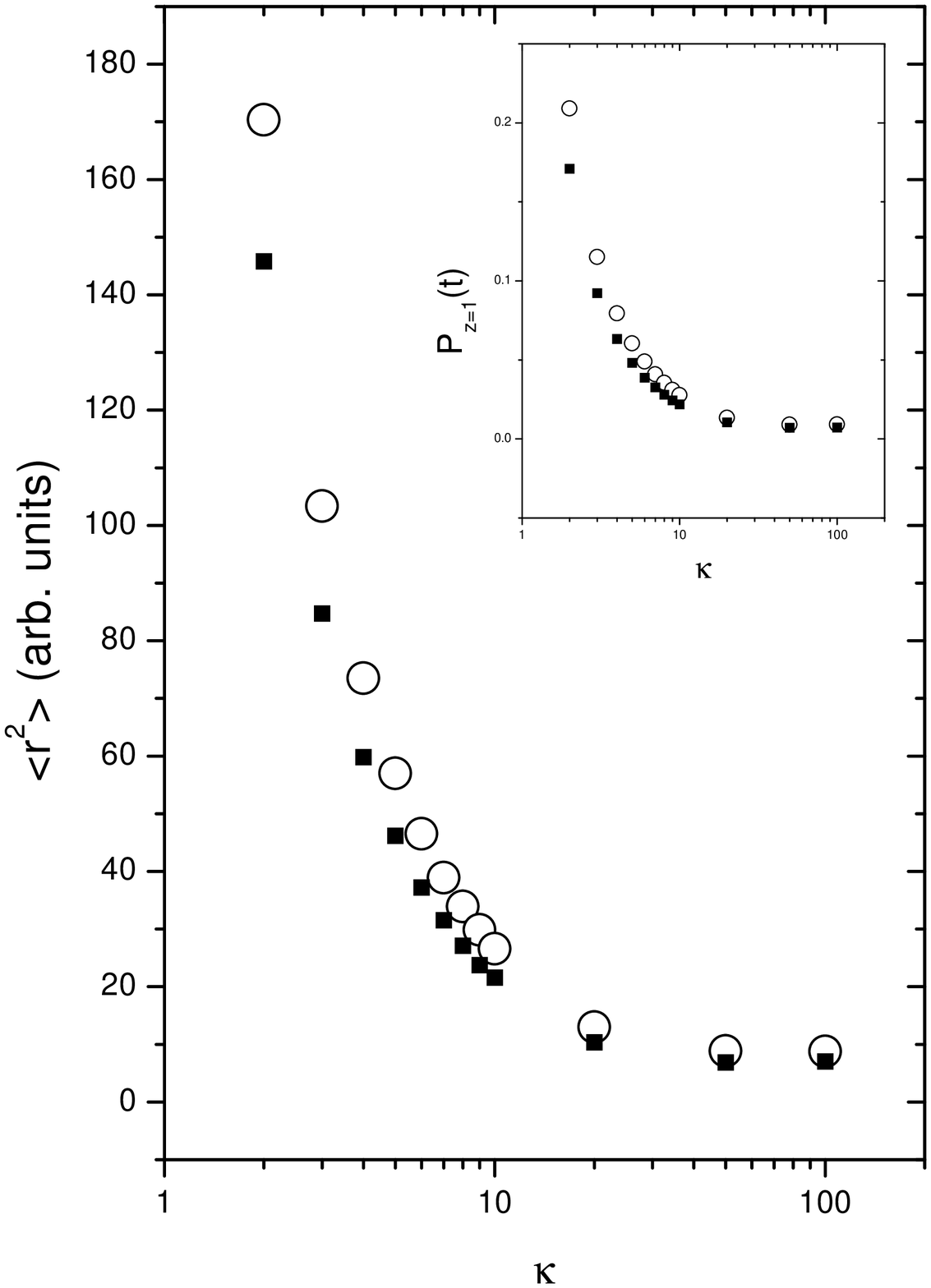}}
\caption{$\langle r^2 \rangle$ vs $\kappa$, The $\bigcirc$
corresponds to a $\delta = 1.5$, $\blacksquare$ to a $\delta = 2$.
Insert: $P_{z=1}(t)$ vs $\kappa$ for these desorption rates.}
\label{f76}
\end{figure}

Figures \ref{f56}, \ref{f66} and \ref{f76} correspond to an
analysis for the asymptotic behavior of the finite system. Figure
\ref{f56} depicts the curve obtained for the $\langle r^2 \rangle$
as a function of $\kappa$ (the number of layers) for three
different and large observational times: $t=1200, 1300, 1500$. The
insert shows that for these times, the system is well inside the
asymptotic region. As can be seen from the figure, there is a
maximum in the motion of the system. In other words, there is an
''optimal" number of layers for which the system can spread more
rapidly. For a larger number of layers, the system converges to an
asymptotic limit. This is reasonable due to the fact that the
finiteness of the system tends to disappear.

Figure \ref{f66} depicts the same behavior as Fig. \ref{f56} but
now we have fixed an observational time ($t=1500$) and we have
used the desorption rate ($\delta$) as parameter. The figure also
shows the maximum on the number of layers again but from this
figure we can see that this maximum moves towards the lower layers
as the desorption rate increases. The insert shows the time
evolution of $\langle r^2 \rangle$ for large values of $\kappa$.
We compare this behavior with the one corresponding to the
infinite case \cite{c6r28}. From this figure we can say that the
finite system behaves, for the used parameters, as an infinite one
when the number of layers is $\kappa > 50$.

If the parameter $\delta$ is increased, the maximum finally
disappears as can be seen in Fig. \ref{f76}. This figure shows two
curves for two different $\delta$ for $t = 1500$. The insert shows
the $P_{z=1}(t)$ for these desorption rates.

\begin{figure}
\centering
\resizebox{.6\columnwidth}{!}{\includegraphics{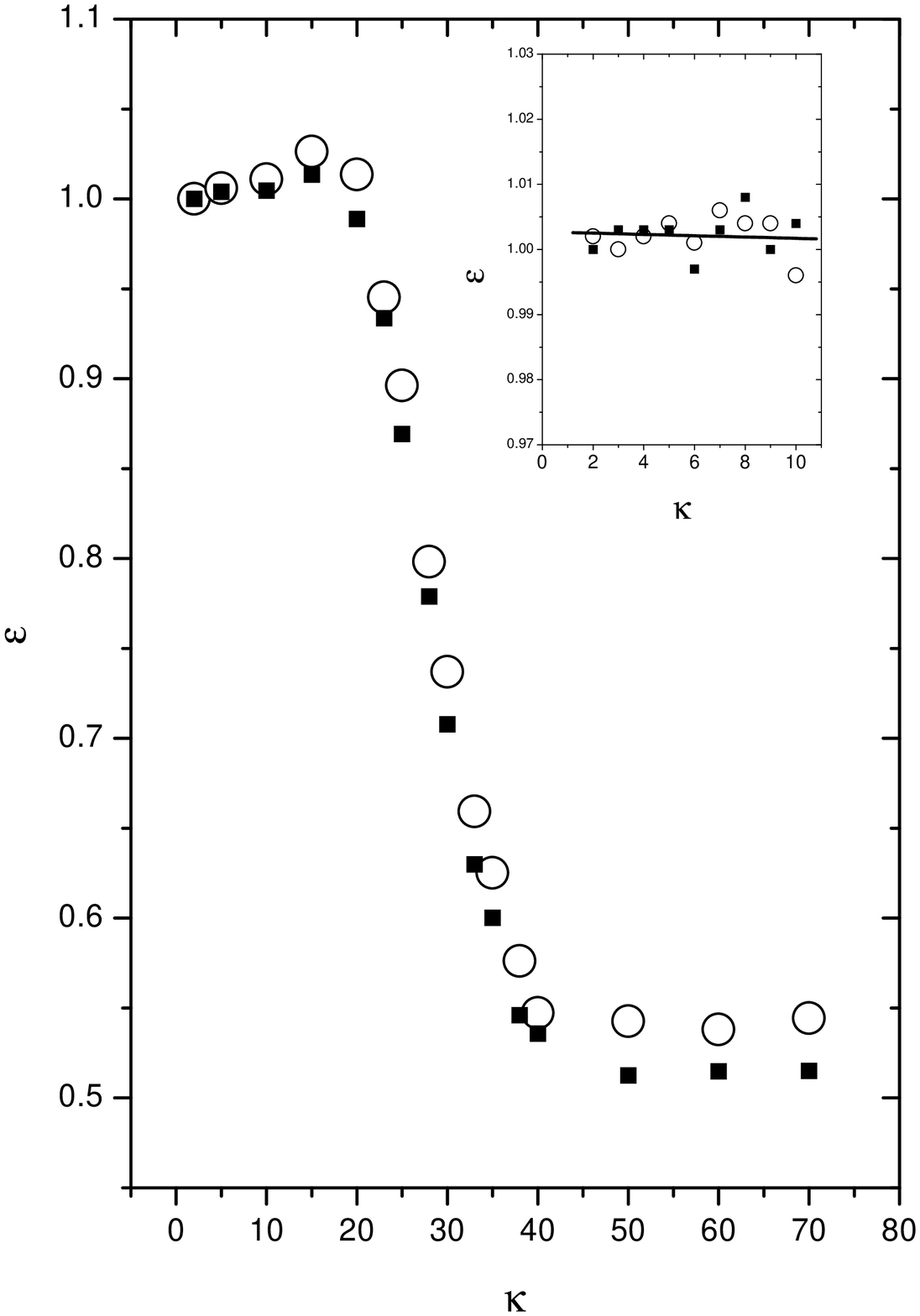}}
\caption{$\epsilon$ vs $\kappa$. Circles correspond to $\delta =
0.5$ and squares to $\delta = 2$. The time evolution was $t=2000$.
Insert: $\epsilon$ vs $\kappa$ for a small number of layers and $t
= 2000$ ($\blacksquare$), $t= 3000$ ($\bigcirc$). The line is a
linear fitting of the data.} \label{f86}
\end{figure}

Finally we show a fitting of $\langle r^2(t) \rangle$ as a
function of $\kappa$ for a large evolution time. For fitting
purposes we have used the following function
\begin{equation}
\label{c5ajuste}
\langle r^2(t) \rangle = C ~ t^\epsilon,
\end{equation}
where $C$ is a constant and $\epsilon$ is the fitting parameter.
In  Fig.  \ref{f86} we depict the results obtained for $\epsilon$
as a function of $\kappa$.

We observe two regions: for $\kappa \leq 20$ the particles
describe a diffusion movement. In particular this behavior was
shown for the bilayer model. The insert of the figure shows a zoom
for a small number of layers, and for two different times. The
linear behavior is apparent. A second region appears for $\kappa
> 20$ where the movement results subdiffusive. For large $\kappa$
the $\epsilon$ parameter approaches an asymptotic value equal to
$0.5$. This situation was predicted \cite{c6r28} for infinite
systems.

\section{Conclusions}

We have studied the evolution of particles diffusing in a volume,
but have analyzed the statistical properties of their evolution on
a surface. The diffusion can be performed on the surface itself or
across the bulk surrounding the surface. In this work we focus on
the particle diffusion due to the movement across the bulk, that
is we have considered the {\it bulk mediated surface diffusion} of
the particles. The main feature of the bulk is its finiteness in
one direction (the axial one). The other directions are unbounded.
This work complements a previous study \cite{c6r28} in which we
have analyzed the particle diffusion in a semi-infinite bulk.

Here we have presented a theoretical model based on a set of
Master Equations which describe the movement of particles over a
simple cubic lattice. This model is general in the sense that we
include both kind of particle movement: on the surface and on the
bulk. We solved this problem by using techniques of the Laplace
and Fourier transformation and exploiting {\it Dyson's formula}.
It is worth remarking here that the model can describe the
evolution of the system everywhere in the bulk, and the evolution
of the system for all time. We particularize to the study over a
single plane, i.e. the surface. We have obtained general solutions
in the Laplace space for the $P_{z=1}(t)$ and the $\langle r^2(t)
\rangle$ for any kind of bulk.

In order to find an analytical solution in space and time, we have
particularized the problem to the bilayer case. The expressions
obtained with this assumption were compared with Monte Carlo
simulations, finding excellent agreement. Moreover we were able to
handle cases with more layers. In these cases, we have obtained
the inverse Laplace transform using numerical methods, and
compared these results with Monte Carlo simulations. The agreement
between both results are excellent again. As indicated above, the
numerical procedure to obtain the inverse Laplace transform is a
reliable tool, and we have shown that we can trust their results
in those cases where analytical results are not accessible.

Finally we have studied the behavior of the spreading in the
asymptotic region as a function of the number of layers of the
system. We observed an ''optimal" number of layers for which
$\langle r^2(t) \rangle$ reaches its maximum value. It is worth
remarking here that this effect occurs in the limit of ''strong
adsorption", that is when the relation $\frac{\delta} {\gamma}$ is
small, and disappears in the limit of ''weak adsorption", that is
when $\frac{\delta}{\gamma}$ is large.

A possible generalization of the present and the previous related
work \cite{c6r28} consists in considering the possibility of non
Markovian dynamics, and study the effect of such dynamics on the
statistical features of the system. This is the subject of further
work.

\vspace{0.25cm}

{\bf Acknowledgments:} The authors thank V. Gr\"unfeld for a
critical reading of the manuscript. HSW acknowledges the partial
support from ANPCyT, Argentine, and thanks the MECyD, Spain, for
an award within the {\it Sabbatical Program for Visiting
Professors}, and to the Universitat de les Illes Balears for the
kind hospitality extended to him.


\end{document}